\documentclass[twocolumn, secnumarabic, amssymb, nobibnotes, superscriptaddress, aps, prb, showpacs]{revtex4-1}
\usepackage{amsmath}
\usepackage{hyperref}
\usepackage{graphicx}
\usepackage{dcolumn}
\usepackage{xfrac}
\newcommand{\ket}[1]{\left| #1 \right>} 
\newcommand{\braket}[2]{\left< #1 \vphantom{#2} \right|
 \left. #2 \vphantom{#1} \right>} 
 
\begin{document}

\title{Measurement of $g$-factor tensor in a quantum dot and disentanglement of exciton spins}

\author{B. J. Witek}
\affiliation{Kavli Institute of Nanoscience, Delft, The Netherlands}
\author{R. W. Heeres}
\affiliation{Kavli Institute of Nanoscience, Delft, The Netherlands}
\author{U. Perinetti}
\affiliation{Kavli Institute of Nanoscience, Delft, The Netherlands}
\author{E. P. A. M. Bakkers}
\affiliation{Kavli Institute of Nanoscience, Delft, The Netherlands}
\affiliation{Philips Research Laboratories, Eindhoven, The Netherlands}
\author{L. P. Kouwenhoven}
\affiliation{Kavli Institute of Nanoscience, Delft, The Netherlands}
\author{V. Zwiller}
\affiliation{Kavli Institute of Nanoscience, Delft, The Netherlands}

\date{\today}

\begin{abstract}
We perform polarization-resolved magneto-optical measurements on single InAsP quantum dots embedded in an InP nanowire. In order to determine all elements of the electron and hole $g$-factor tensors, we measure in magnetic field with different orientations. The results of these measurements are in good agreement with a model based on exchange terms and Zeeman interaction. In our experiment, polarization analysis delivers a powerful tool that not only significantly increases the precision of the measurements, but also enables us to probe the exciton spin state evolution in magnetic fields. We propose a disentangling scheme of heavy-hole exciton spins enabling a measurement of the electron spin $T_2$ time.
\end{abstract}


\maketitle

\section{\label{intro}Introduction}

Carrier spins in semiconductor quantum dots (QDs) have attracted considerable interest due to their potential in quantum information processing based on optical, ultrafast spin manipulation. In recent years, impressive steps toward this goal have been demonstrated: high fidelity spin initialization by optical pumping,\cite{AtatureScience06} coherent population trapping  \cite{XuNature08, Brunner09Science} and coherent spin rotation with picosecond optical pulses.\cite{Press08} In all these cases electron and hole $g$-factors play a crucial role in defining the qubit energy levels. In order to optically address only a single spin state and reduce off-resonant coupling to other states, large $g$-factors are desired. On the other hand, an electron $g$-factor of zero is preferable for coherent photon to spin conversion \cite{Yablonovitch, KosakaNature09, APL2010Kuwahara}.

In bulk semiconductors strong spin-orbit interaction is responsible for relatively large negative electron $g$-factors \cite{Weisbuch77} (e.g. in InAs $g_e=-14.7$ versus free electron $g_e \approx 2$).
Three dimensional confinement in a quantum dot can result in quenching of the orbital angular momentum and hence lead to the modification of $g$-factors.\cite{Kiselev98, Pryor06} The influence of confinement on $g$-factors has been studied in various experiments.\cite{Warburton95, Bjork05, Pryor06, KimPRB09, KleemansPRB09, KlotzAPL10} However, only some of the electron and hole $g$-tensor components were probed [e.g. exciton $g$-factor $g_{X} = (g_{e,z} + g_{h,z})$] providing an incomplete tensor measurement.

Here, we report the results of photoluminescence (PL) measurements in magnetic fields in three different orientations on two differently charged QDs, that reveal all of the components of the electron and hole $g$-factor tensors. Our measurements are polarization - resolved and therefore provide information about magnetic field-induced mixing of quantum states, which was not accessible in previous experiments. \cite{Toft07} The nanowire QDs we use are a promising system for $g$-factor engineering because of the possibility to controllably grow QDs of different sizes and aspect ratios.

\section{\label{model}Model}

In order to describe neutral and charged excitons in a magnetic field, we will utilize the Hamiltonians discussed in detail in Bayer {\it et al.} \cite{Bayer} and van Kesteren {\it et al.}.\cite{Kesteren90} For simplicity only holes that form the top of the valance band, i.e., heavy holes ($J_{h,z}=\pm 3/2$),  are considered here. The electron with a spin $S_{e,z}=\pm 1/2$ ($\uparrow$ or $\downarrow$) and heavy hole with $J_{h,z}=\pm 3/2$ ($\Uparrow$ or $\Downarrow$) can form four exciton states of different total exciton spin projections $J_{z}$:
\begin{equation}
\ket{+1} = \ket{ \downarrow \Uparrow },
\ket{ -1} =\ket{ \uparrow \Downarrow },
\ket{ +2}= \ket{ \uparrow \Uparrow },
\ket{ -2} = \ket{ \downarrow \Downarrow } 
\label{basis}
\end{equation}
The electron and hole spin couple to the external magnetic field via the Zeeman Hamiltonian. By using these four exciton states as the basis, the Zeeman Hamiltonian for a magnetic field $B_z$ oriented along the QD quantization axes (Faraday configuration) can be represented by the matrix \cite{Bayer}:
\begin{align}
H_{B}^{z} &=\frac{\mu_B B_z}{2}
\begin{pmatrix}
g_{X,+ 1}  & 0 & 0 & 0\\
0                 & g_{X,- 1} & 0 & 0\\
0                 & 0                & g_{X,+2} & 0\\
0                 & 0                & 0                & g_{X,-2}
\label{Faraday}
\end{pmatrix}
\end{align}
where $g_{X,\pm 1} =\pm (g_{e,z} + g_{h,z})$ and $g_{X,\pm 2} =\pm (g_{e,z} - g_{h,z})$ are the expressions for bright ($\ket{\pm 1}$) and dark ($\ket{\pm 2}$) exciton $g$-factors in the $z$-direction.

The orientation of the magnetic field in the Faraday configuration matches the QD quantization axis ($z$, growth direction). Therefore the eigenstates of the $H_{B}^{z}$ Hamiltonian coincide with the chosen basis (\ref{basis}).

In the Voigt configuration the magnetic field is applied in the plane of the QD (for simplicity we consider only the $x$ direction) resulting in breaking of the rotational symmetry about the $z$-axis. This leads to the $H_{B}^{x}$ Hamiltonian \cite{Bayer}:
\begin{align}
H_{B}^{x}&=\frac{\mu_B B_x}{2}
\begin{pmatrix}
0  & 0 & g_{e,x}  & g_{h,x}\\
0   & 0 & g_{h,x} & g_{e,x} \\
g_{e,x} & g_{h,x} & 0 & 0\\
g_{h,x} & g_{e,x} & 0 & 0
\label{Voigt}
\end{pmatrix}
\end{align}
The off-diagonal terms account for mixing between bright and dark states. In the Voigt configuration ($B_x$) the resulting eigenvectors are linear superpositions of basis vectors:
\begin{eqnarray}
\ket{X^*_{90^{\circ}}(I)}   &=&\sfrac{1}{2}(\ket{+1}+\ket{-1}+\ket{+2}+\ket{-2}) \nonumber \\
\ket{X^*_{90^{\circ}}(II)} &=&\sfrac{1}{2}(\ket{+1}-\ket{-1}+\ket{+2}-\ket{-2}) \nonumber \\
\ket{X^*_{90^{\circ}}(III)}&=&\sfrac{1}{2}(\ket{+1}-\ket{-1}-\ket{+2}+\ket{-2}) \nonumber \\
\ket{X^*_{90^{\circ}}(IV)}&=&\sfrac{1}{2}(\ket{+1}+\ket{-1}-\ket{+2}-\ket{-2})
\label{basisVoigt}
\end{eqnarray} 
The labels I to IV indicate the eigenstates ordered by increasing energy.

Electron and hole spins do not only interact with an external magnetic field, but also with each other. Exchange interaction couples electron and hole spins in QDs and splits the energy of electron-hole pairs with different spin configurations. The Hamiltonian for the exchange interaction can be written as \cite{Kesteren90, Bayer}:
\begin{align}
H_{exchange}&=\frac{1}{2}
\begin{pmatrix}
+\delta_0 \hfill & +\delta_1 & 0 & 0\\
+\delta_1 & +\delta_0 & 0 & 0\\
0 & 0 & -\delta_0 & +\delta_2\\
0 & 0 & +\delta_2 & -\delta_0
\label{Exchange}
\end{pmatrix}
\end{align}
where $\delta_0$ is the splitting between bright and dark states, $\delta_1$ is often referred in literature as the fine structure splitting (FSS) of a bright exciton and $\delta_2$ is the equivalent splitting for a dark exciton.
Note that the exchange interaction is present for a single electron and hole pair (neutral exciton), and vanishes for more complex exciton molecules that consist of two electrons (total $S_{e,z} = 0$) and/or two holes (total $J_{h,z}=0$). This is the case for a biexciton and singly charged excitons.

\begin{figure}
\includegraphics{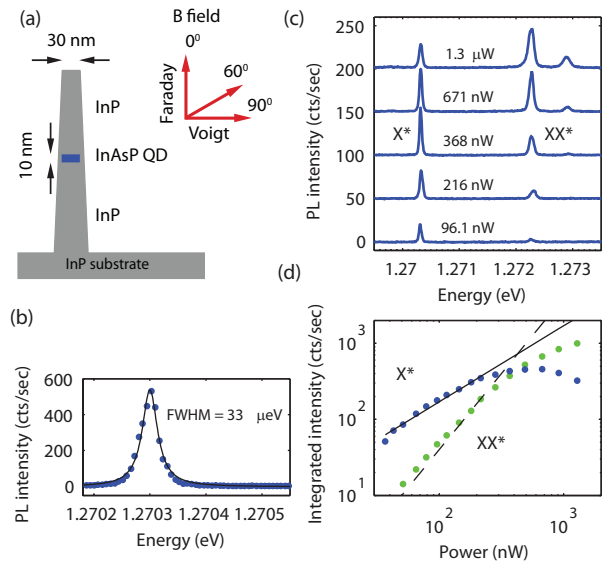}
\caption{\label{general}Structural and optical properties of nanowire quantum dots. (a) Schematic nanowire and magnetic field orientations. (b) PL spectrum of a single exciton recombination. (c) Power dependent spectra taken at 10 K under nonresonant (532 nm) excitation at $B=0$. The two emission peaks are identified as a charged exciton ($X^*$) and charged biexciton ($XX^*$). (d) Integrated intensity of $X^*$ and $XX^*$ transitions versus excitation power. The solid (dashed) line is a guide to the eye for linear (quadratic) power dependence.}
\end{figure}

\section{\label{exp}Experiment}
We studied single InAs$_{0.25}$P$_{0.75}$ wurtzite quantum dots (QDs) embedded in InP wurtzite nanowires grown in the $[1 1 1]$ direction. We performed photoluminescence (PL) measurements with a continuous wave 532 nm excitation laser. The polarization of the QD emission was fully characterized by the tomography measurements using two liquid crystal variable retarders.
Our cryostat ($T\approx 10$ K) with a vector magnet allowed us to vary the direction of the magnetic field in the $x-z$ plane, with $z$ being both the optical axis and the nanowire growth direction. Three magnetic field configurations: Faraday, Voigt and $60^{\circ}$, that are of a particular interest in our experiment, are given in Fig. \ref{general}(a) together with a schematic of a nanowire. A typical QD, with a diameter of 30 nm and height of 10 nm, is surrounded by a thin shell of InP.
Emission linewidths as narrow as 33 $\mu eV$ [Fig. \ref{general}(b)] are clear signatures of the excellent quality of our QDs, which have also demonstrated spin memory in previous studies\cite{NL09Weert}. Power dependent PL spectra presented here [Fig. \ref{general}(c)] belong to a singly charged dot for which the g-factors will be determined in Sec. (\ref{charged}). The PL intensities of the two observed transitions show a linear and a quadratic dependence on excitation power [Fig. \ref{general}(d)] which is consistent with an exciton and biexciton type of recombination. 

\begin{figure*}
\includegraphics{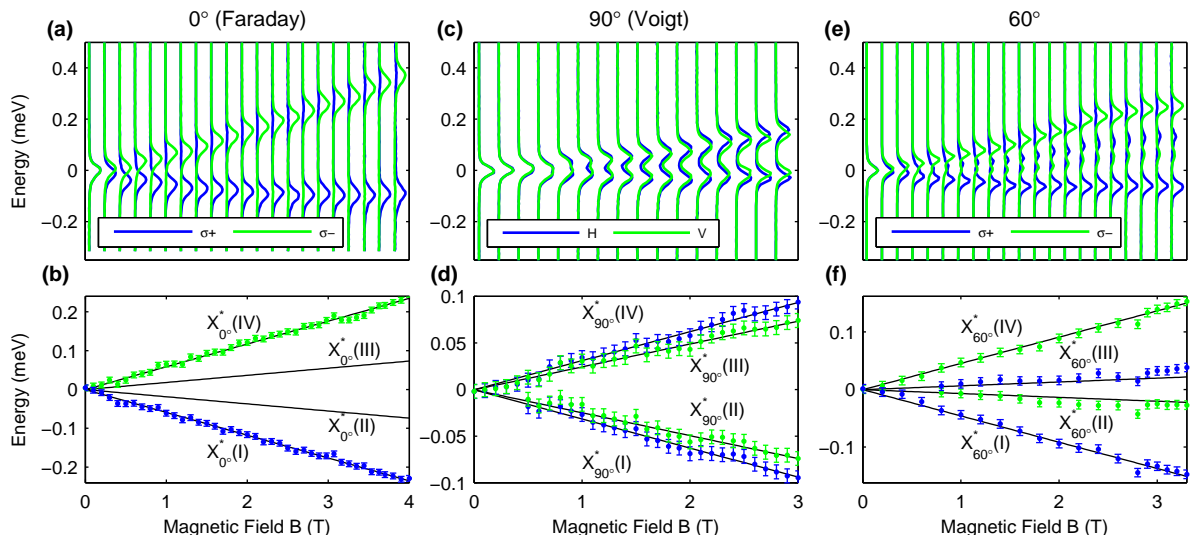}
\caption{\label{charged_fields} Photoluminescence of a charged exciton ($X^*$) in a magnetic field in three different configurations (Faraday, Voigt, $60^{\circ}$). (a, c, e) Polarization resolved spectra. (b, d, f) Energies of the exciton transitions after diamagnetic shift subtraction (data points) fitted with the Zeeman Hamiltonian (lines) for the corresponding magnetic field configuration. The $X^*$ recombination energy at zero magnetic field ($1.2703$ eV) is subtracted. The energy error bar value is $7.4$ $\mu eV$.}
\end{figure*}

\section{\label{charged}Charged Exciton}
We start our discussion with a charged exciton $X^*$ in a magnetic field, since its description is simpler than the neutral exciton case. As mentioned earlier, there is no exchange interaction for $X^*$ and therefore the Zeeman Hamiltonian alone will provide a sufficient model. Although in our experiment we have no means of distinguishing between the positively and negatively charged exciton, the description for both cases is the same. Figure \ref{charged_fields} presents the $X^*$ behavior in three different magnetic field configurations (Faraday, Voigt and $60^{\circ}$). 

The charged exciton PL spectra in the Faraday configuration are given in Fig. \ref{charged_fields}(a). Two exciton states of circular polarization $\sigma ^+$ and $\sigma ^-$ corresponding to the $\ket{+1}$ and $\ket{-1}$ bright states are observed. The peak positions shift with magnetic field due to both Zeeman and diamagnetic effects. After subtracting the quadratic contribution from the diamagnetic shift the exciton state energy versus magnetic field is plotted in Fig. \ref{charged_fields}(b).
The experimental data is fitted with the eigenvalues of the relevant Hamiltonian, in this case $H_B^z$ (\ref{Faraday}), using $g_{X,+1} = g_{e,z}+g_{h,z}$ as the free fitting parameter. Since the two remaining exciton states - dark excitons $X^*_{0^{\circ}}(II)$ and $X^*_{0^{\circ}}(III)$ - are not visible in the Faraday configuration, $g_{X,+2} = g_{e,z}-g_{h,z}$ cannot be extracted from this measurement.

The situation changes in the Voigt configuration due to mixing between states. All four exciton states are present in the PL spectra as evident in Fig. \ref{charged_fields}(c). The transitions are linearly polarized: horizontally $\ket{H} = 1/\sqrt{2}(\ket{-1}+\ket{+1})$ and vertically $\ket{V} = i/\sqrt{2}(\ket{-1}-\ket{+1})$. Interestingly, all four exciton states [$X^*_{90^{\circ}}(I)$, $X^*_{90^{\circ}}(II)$, $X^*_{90^{\circ}}(III)$, and $X^*_{90^{\circ}}(IV)$] have equal measured intensities implying that they must all be equally composed of bright and dark components. These empirical observations are indeed confirmed by analysis of the $H_{B}^x$ eigenvectors from (\ref{basisVoigt}). We find an agreement not only with the fitted energies (Fig. \ref{charged_fields}(d)), but also between observed and predicted brightness and polarization.

From the Faraday and Voigt configurations only in-plane electron $g_{e,x}$ and hole $g_{h,x}$ $g$-factors can be extracted, but the separate values of $g_{e,z}$ and $g_{h,z}$ remain unknown. This missing information is provided by measurements at an intermediate angle ($60^{\circ}$), where some of the features from the Faraday configuration and from the Voigt configuration are combined. Similarly to the Voigt configuration, all four exciton states are observed in the PL spectra, as shown in Fig. \ref{charged_fields}(e).  Nevertheless, the PL intensity of exciton states $X^*_{60^{\circ}}(II)$ and $X^*_{60^{\circ}}(III)$ is on average three times weaker than the PL intensity of states $X^*_{60^{\circ}}(I)$ and $X^*_{60^{\circ}}(IV)$. The transitions are circularly polarized, just like in the case of the Faraday configuration. The fit from Fig. \ref{charged_fields}(f) completes the set of g-factors that is summarized in Table \ref{factors}. Moreover, from the same fit we obtain the expected degree of mixing between dark and bright states. States $X^*_{60^{\circ}}(II)$ and $X^*_{60^{\circ}}(III)$, which were completely dark in the Faraday configuration, now consist of 26\% bright components ($\ket{-1}$ and $\ket{+1}$ respectively). This gain in brightness comes at the expense of states $X^*_{60^{\circ}}(I)$ and $X^*_{60^{\circ}}(IV)$ whose brightness drops to 74\% (compared to 100\% in the Faraday configuration). These predictions match very well with our experimental observations. 

\begin{figure}
\includegraphics{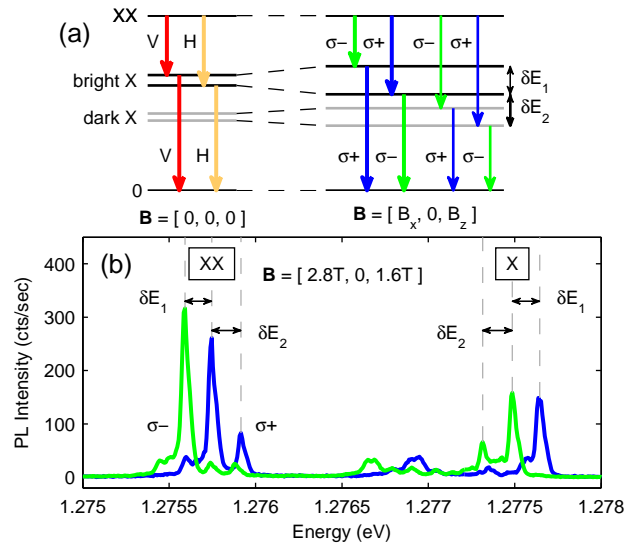}
\caption{\label{neutral60}(a) Schematic of the neutral exciton and biexciton energy levels in a magnetic field at $60^{\circ}$ and (b) corresponding spectrum for $B_{60^{\circ}} = 3.2$ T ({\bf B} = [ $2.8$ T, 0, $1.6$ T ]).}
\end{figure}

\section{\label{neutral}Neutral Exciton}
In the analysis of the neutral exciton $X^0$ in a magnetic field one has to take into account not only the Zeeman Hamiltonian [(\ref{Faraday}) and (\ref{Voigt})], but also the exchange interaction Hamiltonian (\ref{Exchange}). Although the exchange energies are much smaller than $1$ $meV$ and might seem to give only a small correction, the actual effect on the polarization of the eigenstates will prove to be tremendous.

\begin{figure*}
\includegraphics{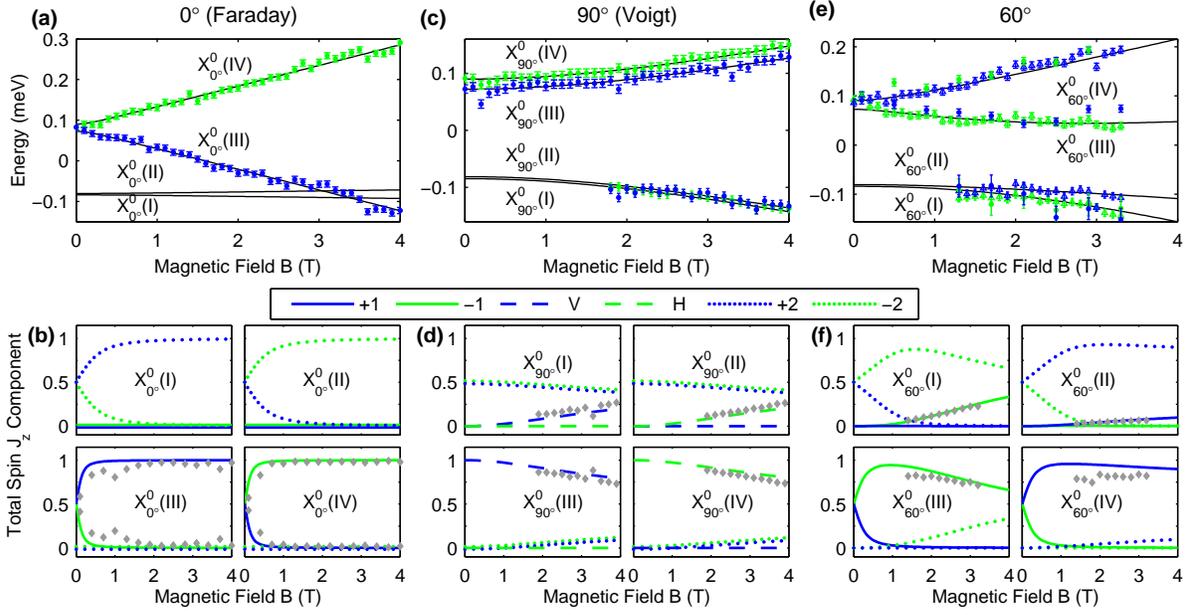}
\caption{\label{neutral_fields}Photoluminescence of a neutral exciton ($X^0$) in a magnetic field in three different configurations (Faraday, Voigt, $60^{\circ}$). (a), (c), (e) Energies of the exciton transitions after diamagnetic shift subtraction (data points) fitted with the Zeeman Hamiltonian (lines) for the corresponding magnetic field configuration. The average of the bright and dark exciton recombination energy at zero magnetic field $1.2775$ eV) is subtracted. The energy error bar value is $7.4$ $\mu eV$. (b ,d , f) Projection of the four exciton states eigenvectors onto the $\ket{+1}$, $\ket{-1}$, $\ket{+2}$, $\ket{-2}$ basis, as well as their linear combination $\ket{H}$ and $\ket{V}$.}
\end{figure*}

Figure \ref{neutral60}(a) presents the schematic of the neutral exciton and biexciton levels in a magnetic field at an intermediate angle (both $x$ and $z$ field components). The biexciton, with a total spin $S_{e,z} = 0$ and $J_{h,z}=0$, experiences no Zeeman effect and no exchange interaction and therefore its recombination energies perfectly mirror those of the exciton transitions. Thus, one can identify the same splittings, for instance $\delta E_1$ and $\delta E_2$ in Fig. \ref{neutral60}(a), in both the biexciton and the exciton emission. We will take advantage of this simple fact and use the biexciton emission to increase the precision of our measurement [especially Fig. \ref{neutral_fields}(e)]. Figure \ref{neutral60} (b) presents the spectrum of a neutral dot in a magnetic field at $60^{\circ}$, where indeed, $\delta E_1$ and $\delta E_2$ have the same magnitude for the exciton and the biexciton.

Figure \ref{neutral_fields} shows the results of measurements on a neutral QD in the same magnetic field configurations as previously discussed for the charged QD. We begin with the Faraday configuration. Two states $X^0_{0^{\circ}}(III)$ and $X^0_{0^{\circ}}(IV)$ are visible and their energies (data points) are given in Fig. \ref{neutral_fields}(a). The sum of the exchange and Zeeman Hamiltonians $H_{exchange}+H_{B}^{z}$ is diagonalized in order to extract the four energy eigenvalues that fit our data (lines). Simultaneously, we also obtain the eigenvectors corresponding to the four exciton states. Each eigenvector, representing one state, can be projected onto the total exciton spin $J_{z}$ basis from (\ref{basis}). All the considered $J_{z}$  components are listed in the legend. Fig. \ref{neutral_fields}(b) consists of four plots, each describing the $J_{z}$ components of a given exciton state. Only states $X^0_{0^{\circ}}(III)$ and $X^0_{0^{\circ}}(IV)$ are bright and hence measurable. At zero magnetic field they both consist equally of $\ket{+1}$ and $\ket{-1}$ resulting in linearly polarized neutral exciton transitions. With increasing magnetic field the transitions evolve toward pure $\sigma^+$ ($J_{z}=\ket{+1}$) and pure $\sigma^-$ ($J_{z}=\ket{-1}$) polarization. The degree of circular polarization was measured for exciton states $X^0_{0^{\circ}}(III)$ and $X^0_{0^{\circ}}(IV)$ and plotted as gray diamonds. The agreement with the predicted curve is very good.

In the Voigt configuration, mixing between dark and bright states becomes sufficiently strong around $B_{x}=2$ T to reveal all four exciton states, whose energies are plotted in Fig. \ref{neutral_fields}(c). In contrast to the charged exciton case (Fig. \ref{charged_fields}(d)), all the exciton states are already split at $B=0$ by the exchange energies ($\delta_0$, $\delta_1$ and $\delta_2$). The solid lines give the fitted eigenvalues of the $H_{exchange}+H_{B}^{x}$ Hamiltonian. As apparent in Fig. \ref{neutral_fields}(d), states $X^0_{90^{\circ}}(I)$ and $X^0_{90^{\circ}}(II)$ start as completely dark at zero magnetic field. With increasing magnetic field they acquire bright components: $\ket{H}$ and $\ket{V}$ respectively, and hence become detectable in PL. This gain in brightness (up to 20 \% at $B_{x}=4$ T) comes at the expense of  $X^0_{90^{\circ}}(III)$ and $X^0_{90^{\circ}}(IV)$ states. These initially purely bright states mix with dark components $\ket{+2}$ and $\ket{-2}$, decreasing the contribution of $\ket{H}$ and $\ket{V}$ to only 80\% at $B_{x}=4$ T.
Based on the results given by our model, we draw the conclusion, that state $X^0_{90^{\circ}}(I)$ couples to $X^0_{90^{\circ}}(III)$ and state $X^0_{90^{\circ}}(II)$ couples to $X^0_{90^{\circ}}(IV)$. The coupled states share the same symmetry, the first pair form an antisymmetric superposition of spins (at $B = 0$  $\ket{X^0_{90^{\circ}}(I)} = 1/\sqrt{2}(\ket{\uparrow \Uparrow}-\ket{\downarrow \Downarrow})$ and $\ket{X^0_{90^{\circ}}(III)}=1/\sqrt{2}(\ket{\downarrow \Uparrow}-\ket{\uparrow \Downarrow})$), whereas the second pair of coupled states is a symmetric superposition of spins. The magnetic field $B_{x}$ is responsible for the precession of the carrier spins around the $x$-axis and therefore couples the states of the same symmetry.

The total brightness of a pair of coupled states can be defined as the sum of their bright components. This brightness per pair is conserved, for instance: $|\braket{V} {X^0_{90^{\circ}}(I)}|^2+|\braket{V} {X^0_{90^{\circ}}(III)}|^2=1$ and $|\braket{H} {X^0_{90^{\circ}}(II)}|^2+|\braket{H} {X^0_{90^{\circ}}(IV)}|^2=1$. Using the above expressions as normalization factors, the contribution of the $\ket{V}$ component to the states $X^0_{90^{\circ}}(I)$ and $X^0_{90^{\circ}}(III)$ (and the $\ket{H}$ component to the states $X^0_{90^{\circ}}(II)$ and $X^0_{90^{\circ}}(IV)$) can be determined from the experiment. The data points (gray diamonds) obtained in this way again follow our predictions with good accuracy.

In the case of the intermediate angle ($60^{\circ}$) in Fig. \ref{neutral_fields}(e), the biexciton recombination energies (mirrored about 0) are represented by triangles, whereas the exciton data are shown by circles. The evolution of the initially bright states, $X^0_{60^{\circ}}(III)$ and $X^0_{60^{\circ}}(IV)$, is equally well reflected by both sets of data points, which confirms the equivalence of $X^0$ and $XX^0$ in our experiment.

We focus now on the analysis of $J_{z}$ components of the exciton states measured in the $60^{\circ}$ configuration. First of all, a striking asymmetry in the exciton states total spin $J_{z}$ composition is immediately recognized in Fig.\ref{neutral_fields}(f). Unlike in the Voigt configuration (Fig. \ref{neutral_fields}(d)), states $X^0_{60^{\circ}}(I)$ and $X^0_{60^{\circ}}(II)$ reach a very different contribution of bright components. At $B_{60^{\circ}}=4$ T it is 34\% of $\ket{+1}$ for state $X^0_{60^{\circ}}(I)$ and only 10\% of $\ket{-1}$ for state $X^0_{60^{\circ}}(II)$. Consequently, bright components of states $X^0_{60^{\circ}}(III)$ and $X^0_{60^{\circ}}(IV)$ evolve unevenly; they drop from 100\% at $B_{60^{\circ}}=0$  to 66\% and 90\% respectively at $B_{60^{\circ}}=4$ T. Following the same procedure as described for the Voigt configuration, we add experimental data points (grey diamonds). There is a good correspondence between the predicted asymmetry and our measurement. 

There is a simple explanation for this phenomenon.  The energy separation between the exciton states sets the strength of the coupling. The gap between states $X^0_{60^{\circ}}(II)$ and $X^0_{60^{\circ}}(IV)$ [Fig. \ref{neutral_fields}(e)] is significantly bigger than for the other pair of states ($X^0_{60^{\circ}}(I)$ and $X^0_{60^{\circ}}(III)$). This results in a much weaker coupling, which is the main reason for the practical difficulties in detecting the extremely weak PL emission from the $X^0_{60^{\circ}}(II)$ state.
The same arguments should also be applied to the Voigt configuration [Fig. \ref{neutral_fields}(c)]. In this case, however, the splittings between the coupled pairs of  states are similar and the asymmetry becomes a negligible effect.

\section{\label{discussion}Discussion}
\begin{table}
\caption{\label{factors} Values of g-factors and exchange energies for both neutral and charged excitons.}
\begin{ruledtabular}
\begin{tabular}{l c c}
 $g$-factor & Neutral Exciton $X^0$ & Charged Exciton $X^*$\\
\hline
$g_{e,z}$   &        $-0.84 \pm 0.02$         &         $-0.70 \pm 0.02$        \\
$g_{h,z}$   &        $-0.92 \pm 0.03$         &         $-1.33 \pm 0.04$        \\
\hline
$|g_{e,x}|$   &        $0.96 \pm 0.02$         &         $1.00 \pm 0.02$        \\
$|g_{h,x}|$   &        $0.04 \pm 0.02$         &         $0.12 \pm 0.01$        \\
\hline \hline
 exchange \\
\hline
$\delta_0 $ &     $163.7 \pm 2.2$ $\mu eV$     &            -          \\
$\delta_1 $ &     $17.7 \pm 2.0 $ $\mu eV$     &            -          \\
$\delta_2 $ &     $3.5 \pm 2.9 $ $\mu eV$      &            -          \\
\hline \hline
\multicolumn{2}{l}{diamagnetic coefficient} \\
\hline
$\gamma_{0^{\circ}}$   &  $11.4 \pm 1.9$ $\mu eV/T^2$  &  $9.7 \pm 0.7$ $\mu eV/T^2$ \\
$\gamma_{60^{\circ}}$  &  $9.5 \pm 1.7$ $\mu eV/T^2$  &  $8.2 \pm 1.0$ $\mu eV/T^2$ \\
$\gamma_{90^{\circ}}$  &  $7.3 \pm 1.9$ $\mu eV/T^2$  &  $7.4 \pm 1.0$ $\mu eV/T^2$ \\
\end{tabular}
\end{ruledtabular}
\end{table}

The complete set of $g$-factors, exchange energies and diamagnetic coefficients obtained from the fits for charged and neutral excitons is summarized in Table \ref{factors}. The $g$-factor components and exchange terms $\delta_0$, $\delta_1$ are determined with high precision. The magnitude of the FSS ($\delta_1$) is confirmed by an additional measurement performed at zero magnetic field as a function of the PL polarization angle ($\delta_1 = 16.1 \pm 2.6$ $\mu eV$). The diamagnetic coefficients $\gamma_{0^{\circ}}$ and $\gamma_{90^{\circ}}$ confirm a stronger confinement along the $z$-direction.

In principle our nanowire QD should exhibit $C_{3v}$ symmetry. \cite{Bester09} However, several factors might lower this. First of all the bottom interface of the QD is sharper than the top one. Second, the randomness of alloying could further reduce the symmetry. In our case, the Hamiltonians [Eq. (\ref{Faraday}), (\ref{Voigt}), (\ref{Exchange})] that are attributed to a symmetry lower than $D_{2d}$ or even no symmetry at all \cite{Bayer} reproduce the experimental results very well.

Performing this type of $g$-factor mapping experiment,\cite{Toft07} we obtain the polarization resolved spectra and therefore access
to the sign of electron and hole $g$-factors along the $z$ direction. Strong confinement responsible for orbital angular momentum quenching \cite{Pryor06} pushes the exciton $g$-factor to positive values, as reported for a similar InAs/InP self-assembled QDs system \cite{KimPRB09, KleemansPRB09} ($g_X$ as high as 1.25). In our case, however, the exciton $g$-factor is found to be negative ($g_X = g_{e,z}+g_{h,z} $ is $-1.76$ and $-2.03$), implying a weaker confinement. Indeed, the average height of the NW QD is larger than for self-assembled QDs.
In case of the in-plane $g$-factors it is not possible to tell if they are positive or negative, since the polarization of exciton states in the Voigt configuration is insensitive to their sign.

The in-plane hole $g$-factor $g_{h,x}$ is almost zero, which resembles the situation in quantum wells \cite{Marie99, Glasberg99}. In theory the heavy hole in-plane $g$-factor is almost negligible and mostly determined by a Luttinger $q$ parameter \cite{Marie99}, $g_{hh,x}\approx3q$, whereas the light hole $g$-factor takes larger nonzero values \cite{Kiselev01}. 
For bulk InAs and InP, Luttinger parameters are 0.04 and 0.02 respectively \cite{Dymnikov09} leading to an estimated $g_{hh,x}\approx0.09$. The experimental result for nanowire QDs deviates slightly from this approximation, which is not surprising taking into account the 3D confinement. The charged exciton in-plane hole g-factor ($g_{h,x}=0.12 \pm 0.01$) is larger than the neutral one ($g_{h,x}=0.04 \pm 0.02$), and the same trend was reported for interfacial QDs in GaAs QWs \cite{Toft07}. Still this value is not large enough to imply heavy hole - light hole coupling. Although in our analysis we completely neglect light holes, we still obtain a very precise description of the experiment, which confirms the validity of our assumption.

The exciton spin behavior is substantially different for the charged and neutral exciton. As apparent from the charged exciton spectra in Fig. \ref{charged_fields} the polarization and relative intensity of the transitions is independent of the magnetic field magnitude, implying that the mixing between bright and dark states is constant. Dark states become visible immediately in the nonzero transverse magnetic field, which is crucial in experiments involving a three-level lambda system formed by charged exciton states in the Voigt configuration \cite{XuNature08, Brunner09Science}. On the other hand, for the neutral exciton the strength of bright-dark state mixing increases with magnetic field (Fig. \ref{neutral_fields}). The separation $\delta_0$ between the bright and dark states prevents the immediate coupling. 

\begin{figure}
\includegraphics{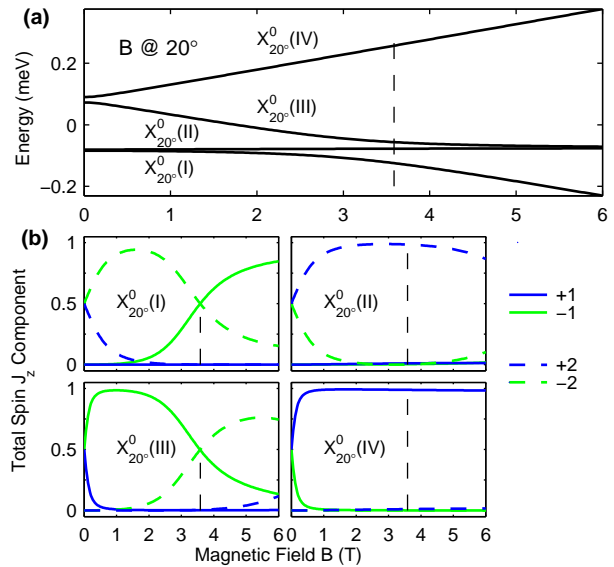}
\caption{\label{sym}Simulation of neutral exciton states in a magnetic field at $20^{\circ}$. The exchange energies and $g$-factors are the experimental values. (a) Energy shift due to the Zeeman effect and (b) the projection of the four exciton states eigenvectors onto the $\ket{+1}$, $\ket{-1}$, $\ket{+2}$, $\ket{-2}$ basis.}
\end{figure}

\begin{figure}
\includegraphics{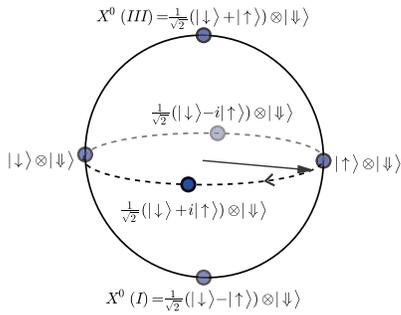}
\caption{\label{bloch} Coherent time evolution of the electron spin disentangled from the heavy-hole spin.}
\end{figure}

Our studies have demonstrated that the coupling strength between exciton states can be tuned by a careful choice of the magnetic field angle and magnitude. This opens the possibility of engineering any superposition of exciton spin states at will. One particular example is illustrated in Fig. \ref{sym}, where the behavior of a neutral exciton in a magnetic field at $20^{\circ}$ is simulated. We took the experimental values of exchange energies and $g$-factors listed in Table \ref{factors}. Fig. \ref{sym}(a) plots the energies of exciton states in a magnetic field up to 6 T. We observe the anticrossing of states $X^0_{20^{\circ}}(I)$ and $X^0_{20^{\circ}}(III)$. At the anticrossing, at approximately $B_{20^{\circ}} = 3.6$ T, the exciton spin states take a special form. As evident from Fig. \ref{sym}(b) states $X^0_{20^{\circ}}(I)$ and $X^0_{20^{\circ}}(III)$ are equally composed of dark $\ket{-2}$ and bright $\ket{-1}$ components. At the same time states $X^0(II)$ and $X^0(IV)$ do not mix and stay completely dark $\ket{+2}$ or completely bright $\ket{+1}$. We can write the corresponding spin states at $B_{20^{\circ}} = 3.6$ T as follows:
\begin{eqnarray}
\ket{X^0_{20^{\circ}}(I)}_{3.6T} &=& \frac{1}{\sqrt{2}}(\ket{ \downarrow \Downarrow }- \ket{\uparrow \Downarrow })=\frac{1}{\sqrt{2}}(\ket{\downarrow}-\ket{\uparrow})\otimes \ket{\Downarrow} \nonumber \\
\ket{ X^0_{20^{\circ}}(II)}_{3.6T} &=&\ket{ \uparrow \Uparrow } \nonumber \\
\ket{X^0_{20^{\circ}}(III)}_{3.6T} &=& \frac{1}{\sqrt{2}}(\ket{ \downarrow \Downarrow }+ \ket{\uparrow \Downarrow })=\frac{1}{\sqrt{2}}(\ket{\downarrow}+\ket{\uparrow})\otimes \ket{\Downarrow} \nonumber \\
\ket{ X^0_{20^{\circ}}(IV)}_{3.6T} &=&\ket{ \downarrow \Uparrow }
\label{magic}
\end{eqnarray}
The hole spin $\ket{\Downarrow}$ can be factored out from states $X^0_{20^{\circ}}(I)$ and $X^0_{20^{\circ}}(III)$ leaving the superposition of electron spin states. Note that this is possible only because of a very small in-plane hole $g$-factor, which ensures that the hole spin stays insensitive to $x$-components of the magnetic field. Based on these properties we propose a scheme that enables measuring the electron coherence time $T_2$ in a similar fashion to the experiment by Kroutvar {\it et al.} \cite{Nature04Finley} on electron spin relaxation time $T_1$. A left handed circularly polarized pump pulse ($\sigma^-$) can create a superposition $\ket{\uparrow \Downarrow} = 1/\sqrt{2}( \ket{X^0_{20^{\circ}}(I)}_{3.6T}-\ket{X^0_{20}(III)}_{3.6T})$ that precesses in time at a frequency given by the difference between the eigenenergies $\Delta E$. This situation is illustrated in Fig. \ref{bloch}, where the photocreated state precesses in the equator plane of the Bloch sphere. The most crucial feature in this experiment is that we can factor out and therefore disentangle the hole state and only consider the electron superposition  $1/\sqrt{2}( \ket{ \uparrow}-e^{i\Delta Et/\hbar}\ket{\downarrow})\otimes\ket{\Downarrow}$. Under the application of external electric field one can remove the disentangled hole from the QD without any harm to the coherence of the electron spin superposition. After a certain delay time a hole can be brought back to the QD. This will result in a photon emission, whose polarization should exhibit quantum beats. The envelope of the beats is set by the electron $T_2$. This suggested method will enable the $T_2$ measurement in time resolved PL. Other techniques of probing $T_{2}^*$, such as time resolved Faraday \cite{Gupta01Science} and Kerr \cite{Mikkelsen07Nature, Berezovsky08Science, Kosaka08} rotation are based on transmission and reflection measurements respectively.

\section{\label{conclusion}Conclusion}
In conclusion, we have presented a set of magneto - optical measurements that leads to the precise determination of $g$-factor tensor components for the electron and hole.
In addition, the possibility of polarization analysis has given us a tool to probe the exciton spin response to the external magnetic field in any configuration. Our model has proven to provide a complete and self-consistent description of all the observed experimental effects in magnetic fields, from the evolution of the energy of exciton states to the prediction of their spin eigenstates. We have proposed a scheme of disentangling heavy-hole exciton spins that opens a way of measuring electron spin coherence $T_2$ decoupled from a hole.

\begin{acknowledgments}
We would like to acknowledge Craig Pryor for helpful discussions oncerning the NW QD system symmetry. This work was supported by the European NanoSci ERA-net project, the Dutch Organization for Fundamental Research on Matter (FOM) and The
Netherlands Organization for Scientific Research (NWO).
\end{acknowledgments}


\begin{thebibliography}{29}%
\makeatletter
\providecommand \@ifxundefined [1]{%
 \@ifx{#1\undefined}
}%
\providecommand \@ifnum [1]{%
 \ifnum #1\expandafter \@firstoftwo
 \else \expandafter \@secondoftwo
 \fi
}%
\providecommand \@ifx [1]{%
 \ifx #1\expandafter \@firstoftwo
 \else \expandafter \@secondoftwo
 \fi
}%
\providecommand \natexlab [1]{#1}%
\providecommand \enquote  [1]{``#1''}%
\providecommand \bibnamefont  [1]{#1}%
\providecommand \bibfnamefont [1]{#1}%
\providecommand \citenamefont [1]{#1}%
\providecommand \href@noop [0]{\@secondoftwo}%
\providecommand \href [0]{\begingroup \@sanitize@url \@href}%
\providecommand \@href[1]{\@@startlink{#1}\@@href}%
\providecommand \@@href[1]{\endgroup#1\@@endlink}%
\providecommand \@sanitize@url [0]{\catcode `\\12\catcode `\$12\catcode
  `\&12\catcode `\#12\catcode `\^12\catcode `\_12\catcode `\%12\relax}%
\providecommand \@@startlink[1]{}%
\providecommand \@@endlink[0]{}%
\providecommand \url  [0]{\begingroup\@sanitize@url \@url }%
\providecommand \@url [1]{\endgroup\@href {#1}{\urlprefix }}%
\providecommand \urlprefix  [0]{URL }%
\providecommand \Eprint [0]{\href }%
\providecommand \doibase [0]{http://dx.doi.org/}%
\providecommand \selectlanguage [0]{\@gobble}%
\providecommand \bibinfo  [0]{\@secondoftwo}%
\providecommand \bibfield  [0]{\@secondoftwo}%
\providecommand \translation [1]{[#1]}%
\providecommand \BibitemOpen [0]{}%
\providecommand \bibitemStop [0]{}%
\providecommand \bibitemNoStop [0]{.\EOS\space}%
\providecommand \EOS [0]{\spacefactor3000\relax}%
\providecommand \BibitemShut  [1]{\csname bibitem#1\endcsname}%
\let\auto@bib@innerbib\@empty
\bibitem [{\citenamefont {Atat\"{u}re}\ \emph {et~al.}(2006)\citenamefont
  {Atat\"{u}re}, \citenamefont {Dreiser}, \citenamefont {Badolato},
  \citenamefont {H\"{o}gele}, \citenamefont {Karrai},\ and\ \citenamefont
  {Imamoglu}}]{AtatureScience06}%
  \BibitemOpen
  \bibfield  {author} {\bibinfo {author} {\bibfnamefont {M.}~\bibnamefont
  {Atat\"{u}re}}, \bibinfo {author} {\bibfnamefont {J.}~\bibnamefont
  {Dreiser}}, \bibinfo {author} {\bibfnamefont {A.}~\bibnamefont {Badolato}},
  \bibinfo {author} {\bibfnamefont {A.}~\bibnamefont {H\"{o}gele}}, \bibinfo
  {author} {\bibfnamefont {K.}~\bibnamefont {Karrai}}, \ and\ \bibinfo {author}
  {\bibfnamefont {A.}~\bibnamefont {Imamoglu}},\ }\href {\doibase
  10.1126/science.1126074} {\bibfield  {journal} {\bibinfo  {journal}
  {Science}\ }\textbf {\bibinfo {volume} {312}},\ \bibinfo {pages} {551}
  (\bibinfo {year} {2006})}\BibitemShut {NoStop}%
\bibitem [{\citenamefont {Xu}\ \emph {et~al.}(2008)\citenamefont {Xu},
  \citenamefont {Sun}, \citenamefont {Berman}, \citenamefont {Steel},
  \citenamefont {Bracker}, \citenamefont {Gammon},\ and\ \citenamefont
  {Sham}}]{XuNature08}%
  \BibitemOpen
  \bibfield  {author} {\bibinfo {author} {\bibfnamefont {X.}~\bibnamefont
  {Xu}}, \bibinfo {author} {\bibfnamefont {B.}~\bibnamefont {Sun}}, \bibinfo
  {author} {\bibfnamefont {P.}~\bibnamefont {Berman}}, \bibinfo {author}
  {\bibfnamefont {D.}~\bibnamefont {Steel}}, \bibinfo {author} {\bibfnamefont
  {A.}~\bibnamefont {Bracker}}, \bibinfo {author} {\bibfnamefont
  {D.}~\bibnamefont {Gammon}}, \ and\ \bibinfo {author} {\bibfnamefont
  {L.}~\bibnamefont {Sham}},\ }\href@noop {} {\bibfield  {journal} {\bibinfo
  {journal} {Nature Physics}\ }\textbf {\bibinfo {volume} {4}},\ \bibinfo
  {pages} {692} (\bibinfo {year} {2008})}\BibitemShut {NoStop}%
\bibitem [{\citenamefont {Brunner}\ \emph {et~al.}(2009)\citenamefont
  {Brunner}, \citenamefont {Gerardot}, \citenamefont {Dalgarno}, \citenamefont
  {W{\\"u}st}, \citenamefont {Karrai}, \citenamefont {Stoltz}, \citenamefont
  {Petroff},\ and\ \citenamefont {Warburton}}]{Brunner09Science}%
  \BibitemOpen
  \bibfield  {author} {\bibinfo {author} {\bibfnamefont {D.}~\bibnamefont
  {Brunner}}, \bibinfo {author} {\bibfnamefont {B.}~\bibnamefont {Gerardot}},
  \bibinfo {author} {\bibfnamefont {P.}~\bibnamefont {Dalgarno}}, \bibinfo
  {author} {\bibfnamefont {G.}~\bibnamefont {W{\\"u}st}}, \bibinfo {author}
  {\bibfnamefont {K.}~\bibnamefont {Karrai}}, \bibinfo {author} {\bibfnamefont
  {N.}~\bibnamefont {Stoltz}}, \bibinfo {author} {\bibfnamefont
  {P.}~\bibnamefont {Petroff}}, \ and\ \bibinfo {author} {\bibfnamefont
  {R.}~\bibnamefont {Warburton}},\ }\href@noop {} {\bibfield  {journal}
  {\bibinfo  {journal} {Science}\ }\textbf {\bibinfo {volume} {325}},\ \bibinfo
  {pages} {70} (\bibinfo {year} {2009})}\BibitemShut {NoStop}%
\bibitem [{\citenamefont {Press}\ \emph {et~al.}(2008)\citenamefont {Press},
  \citenamefont {Ladd}, \citenamefont {Zhang},\ and\ \citenamefont
  {Yamamoto}}]{Press08}%
  \BibitemOpen
  \bibfield  {author} {\bibinfo {author} {\bibfnamefont {D.}~\bibnamefont
  {Press}}, \bibinfo {author} {\bibfnamefont {T.}~\bibnamefont {Ladd}},
  \bibinfo {author} {\bibfnamefont {B.}~\bibnamefont {Zhang}}, \ and\ \bibinfo
  {author} {\bibfnamefont {Y.}~\bibnamefont {Yamamoto}},\ }\href@noop {}
  {\bibfield  {journal} {\bibinfo  {journal} {Nature}\ }\textbf {\bibinfo
  {volume} {456}},\ \bibinfo {pages} {218} (\bibinfo {year}
  {2008})}\BibitemShut {NoStop}%
\bibitem [{\citenamefont {Vrijen}\ and\ \citenamefont
  {Yablonovitch}(2001)}]{Yablonovitch}%
  \BibitemOpen
  \bibfield  {author} {\bibinfo {author} {\bibfnamefont {R.}~\bibnamefont
  {Vrijen}}\ and\ \bibinfo {author} {\bibfnamefont {E.}~\bibnamefont
  {Yablonovitch}},\ }\href@noop {} {\bibfield  {journal} {\bibinfo  {journal}
  {Physica E: Low-dimensional Systems and Nanostructures}\ }\textbf {\bibinfo
  {volume} {10}},\ \bibinfo {pages} {569} (\bibinfo {year} {2001})}\BibitemShut
  {NoStop}%
\bibitem [{\citenamefont {Kosaka}\ \emph {et~al.}(2009)\citenamefont {Kosaka},
  \citenamefont {Inagaki}, \citenamefont {Rikitake}, \citenamefont {Imamura},
  \citenamefont {Mitsumori},\ and\ \citenamefont {Edamatsu}}]{KosakaNature09}%
  \BibitemOpen
  \bibfield  {author} {\bibinfo {author} {\bibfnamefont {H.}~\bibnamefont
  {Kosaka}}, \bibinfo {author} {\bibfnamefont {T.}~\bibnamefont {Inagaki}},
  \bibinfo {author} {\bibfnamefont {Y.}~\bibnamefont {Rikitake}}, \bibinfo
  {author} {\bibfnamefont {H.}~\bibnamefont {Imamura}}, \bibinfo {author}
  {\bibfnamefont {Y.}~\bibnamefont {Mitsumori}}, \ and\ \bibinfo {author}
  {\bibfnamefont {K.}~\bibnamefont {Edamatsu}},\ }\href@noop {} {\bibfield
  {journal} {\bibinfo  {journal} {Nature}\ }\textbf {\bibinfo {volume} {457}},\
  \bibinfo {pages} {702} (\bibinfo {year} {2009})}\BibitemShut {NoStop}%
\bibitem [{\citenamefont {Kuwahara}\ \emph {et~al.}(2010)\citenamefont
  {Kuwahara}, \citenamefont {Kutsuwa}, \citenamefont {Ono},\ and\ \citenamefont
  {Kosaka}}]{APL2010Kuwahara}%
  \BibitemOpen
  \bibfield  {author} {\bibinfo {author} {\bibfnamefont {M.}~\bibnamefont
  {Kuwahara}}, \bibinfo {author} {\bibfnamefont {T.}~\bibnamefont {Kutsuwa}},
  \bibinfo {author} {\bibfnamefont {K.}~\bibnamefont {Ono}}, \ and\ \bibinfo
  {author} {\bibfnamefont {H.}~\bibnamefont {Kosaka}},\ }\href@noop {}
  {\bibfield  {journal} {\bibinfo  {journal} {Applied Physics Letters}\
  }\textbf {\bibinfo {volume} {96}},\ \bibinfo {pages} {163107} (\bibinfo
  {year} {2010})}\BibitemShut {NoStop}%
\bibitem [{\citenamefont {Weisbuch}\ and\ \citenamefont
  {Hermann}(1977)}]{Weisbuch77}%
  \BibitemOpen
  \bibfield  {author} {\bibinfo {author} {\bibfnamefont {C.}~\bibnamefont
  {Weisbuch}}\ and\ \bibinfo {author} {\bibfnamefont {C.}~\bibnamefont
  {Hermann}},\ }\href {\doibase 10.1103/PhysRevB.15.816} {\bibfield  {journal}
  {\bibinfo  {journal} {Phys. Rev. B}\ }\textbf {\bibinfo {volume} {15}},\
  \bibinfo {pages} {816} (\bibinfo {year} {1977})}\BibitemShut {NoStop}%
\bibitem [{\citenamefont {Kiselev}\ \emph {et~al.}(1998)\citenamefont
  {Kiselev}, \citenamefont {Ivchenko},\ and\ \citenamefont
  {R\"ossler}}]{Kiselev98}%
  \BibitemOpen
  \bibfield  {author} {\bibinfo {author} {\bibfnamefont {A.~A.}\ \bibnamefont
  {Kiselev}}, \bibinfo {author} {\bibfnamefont {E.~L.}\ \bibnamefont
  {Ivchenko}}, \ and\ \bibinfo {author} {\bibfnamefont {U.}~\bibnamefont
  {R\"ossler}},\ }\href {\doibase 10.1103/PhysRevB.58.16353} {\bibfield
  {journal} {\bibinfo  {journal} {Phys. Rev. B}\ }\textbf {\bibinfo {volume}
  {58}},\ \bibinfo {pages} {16353} (\bibinfo {year} {1998})}\BibitemShut
  {NoStop}%
\bibitem [{\citenamefont {Pryor}\ and\ \citenamefont
  {Flatt\'e}(2006)}]{Pryor06}%
  \BibitemOpen
  \bibfield  {author} {\bibinfo {author} {\bibfnamefont {C.~E.}\ \bibnamefont
  {Pryor}}\ and\ \bibinfo {author} {\bibfnamefont {M.~E.}\ \bibnamefont
  {Flatt\'e}},\ }\href {\doibase 10.1103/PhysRevLett.96.026804} {\bibfield
  {journal} {\bibinfo  {journal} {Phys. Rev. Lett.}\ }\textbf {\bibinfo
  {volume} {96}},\ \bibinfo {pages} {026804} (\bibinfo {year}
  {2006})}\BibitemShut {NoStop}%
\bibitem [{\citenamefont {Traynor}\ \emph {et~al.}(1995)\citenamefont
  {Traynor}, \citenamefont {Harley},\ and\ \citenamefont
  {Warburton}}]{Warburton95}%
  \BibitemOpen
  \bibfield  {author} {\bibinfo {author} {\bibfnamefont {N.~J.}\ \bibnamefont
  {Traynor}}, \bibinfo {author} {\bibfnamefont {R.~T.}\ \bibnamefont {Harley}},
  \ and\ \bibinfo {author} {\bibfnamefont {R.~J.}\ \bibnamefont {Warburton}},\
  }\href {\doibase 10.1103/PhysRevB.51.7361} {\bibfield  {journal} {\bibinfo
  {journal} {Physical Review B}\ }\textbf {\bibinfo {volume} {51}},\ \bibinfo
  {pages} {7361} (\bibinfo {year} {1995})}\BibitemShut {NoStop}%
\bibitem [{\citenamefont {Bj\"ork}\ \emph {et~al.}(2005)\citenamefont
  {Bj\"ork}, \citenamefont {Fuhrer}, \citenamefont {Hansen}, \citenamefont
  {Larsson}, \citenamefont {Fr\"oberg},\ and\ \citenamefont
  {Samuelson}}]{Bjork05}%
  \BibitemOpen
  \bibfield  {author} {\bibinfo {author} {\bibfnamefont {M.~T.}\ \bibnamefont
  {Bj\"ork}}, \bibinfo {author} {\bibfnamefont {A.}~\bibnamefont {Fuhrer}},
  \bibinfo {author} {\bibfnamefont {A.~E.}\ \bibnamefont {Hansen}}, \bibinfo
  {author} {\bibfnamefont {M.~W.}\ \bibnamefont {Larsson}}, \bibinfo {author}
  {\bibfnamefont {L.~E.}\ \bibnamefont {Fr\"oberg}}, \ and\ \bibinfo {author}
  {\bibfnamefont {L.}~\bibnamefont {Samuelson}},\ }\href {\doibase
  10.1103/PhysRevB.72.201307} {\bibfield  {journal} {\bibinfo  {journal} {Phys.
  Rev. B}\ }\textbf {\bibinfo {volume} {72}},\ \bibinfo {pages} {201307}
  (\bibinfo {year} {2005})}\BibitemShut {NoStop}%
\bibitem [{\citenamefont {Kim}\ \emph {et~al.}(2009)\citenamefont {Kim},
  \citenamefont {Sheng}, \citenamefont {Poole}, \citenamefont {Dalacu},
  \citenamefont {Lefebvre}, \citenamefont {Lapointe}, \citenamefont {Reimer},
  \citenamefont {Aers},\ and\ \citenamefont {Williams}}]{KimPRB09}%
  \BibitemOpen
  \bibfield  {author} {\bibinfo {author} {\bibfnamefont {D.}~\bibnamefont
  {Kim}}, \bibinfo {author} {\bibfnamefont {W.}~\bibnamefont {Sheng}}, \bibinfo
  {author} {\bibfnamefont {P.~J.}\ \bibnamefont {Poole}}, \bibinfo {author}
  {\bibfnamefont {D.}~\bibnamefont {Dalacu}}, \bibinfo {author} {\bibfnamefont
  {J.}~\bibnamefont {Lefebvre}}, \bibinfo {author} {\bibfnamefont
  {J.}~\bibnamefont {Lapointe}}, \bibinfo {author} {\bibfnamefont {M.~E.}\
  \bibnamefont {Reimer}}, \bibinfo {author} {\bibfnamefont {G.~C.}\
  \bibnamefont {Aers}}, \ and\ \bibinfo {author} {\bibfnamefont {R.~L.}\
  \bibnamefont {Williams}},\ }\href {\doibase 10.1103/PhysRevB.79.045310}
  {\bibfield  {journal} {\bibinfo  {journal} {Physical Review B}\ }\textbf
  {\bibinfo {volume} {79}},\ \bibinfo {pages} {045310} (\bibinfo {year}
  {2009})}\BibitemShut {NoStop}%
\bibitem [{\citenamefont {Kleemans}\ \emph {et~al.}(2009)\citenamefont
  {Kleemans}, \citenamefont {van Bree}, \citenamefont {Bozkurt}, \citenamefont
  {van Veldhoven}, \citenamefont {Nouwens}, \citenamefont {N\"otzel},
  \citenamefont {Silov}, \citenamefont {Koenraad},\ and\ \citenamefont
  {Flatt\'e}}]{KleemansPRB09}%
  \BibitemOpen
  \bibfield  {author} {\bibinfo {author} {\bibfnamefont {N.~A. J.~M.}\
  \bibnamefont {Kleemans}}, \bibinfo {author} {\bibfnamefont {J.}~\bibnamefont
  {van Bree}}, \bibinfo {author} {\bibfnamefont {M.}~\bibnamefont {Bozkurt}},
  \bibinfo {author} {\bibfnamefont {P.~J.}\ \bibnamefont {van Veldhoven}},
  \bibinfo {author} {\bibfnamefont {P.~A.}\ \bibnamefont {Nouwens}}, \bibinfo
  {author} {\bibfnamefont {R.}~\bibnamefont {N\"otzel}}, \bibinfo {author}
  {\bibfnamefont {A.~Y.}\ \bibnamefont {Silov}}, \bibinfo {author}
  {\bibfnamefont {P.~M.}\ \bibnamefont {Koenraad}}, \ and\ \bibinfo {author}
  {\bibfnamefont {M.~E.}\ \bibnamefont {Flatt\'e}},\ }\href {\doibase
  10.1103/PhysRevB.79.045311} {\bibfield  {journal} {\bibinfo  {journal} {Phys.
  Rev. B}\ }\textbf {\bibinfo {volume} {79}},\ \bibinfo {pages} {045311}
  (\bibinfo {year} {2009})}\BibitemShut {NoStop}%
\bibitem [{\citenamefont {Klotz}\ \emph {et~al.}(2010)\citenamefont {Klotz},
  \citenamefont {Jovanov}, \citenamefont {Kierig}, \citenamefont {Clark},
  \citenamefont {Rudolph}, \citenamefont {Heiss}, \citenamefont {Bichler},
  \citenamefont {Abstreiter}, \citenamefont {Brandt},\ and\ \citenamefont
  {Finley}}]{KlotzAPL10}%
  \BibitemOpen
  \bibfield  {author} {\bibinfo {author} {\bibfnamefont {F.}~\bibnamefont
  {Klotz}}, \bibinfo {author} {\bibfnamefont {V.}~\bibnamefont {Jovanov}},
  \bibinfo {author} {\bibfnamefont {J.}~\bibnamefont {Kierig}}, \bibinfo
  {author} {\bibfnamefont {E.~C.}\ \bibnamefont {Clark}}, \bibinfo {author}
  {\bibfnamefont {D.}~\bibnamefont {Rudolph}}, \bibinfo {author} {\bibfnamefont
  {D.}~\bibnamefont {Heiss}}, \bibinfo {author} {\bibfnamefont
  {M.}~\bibnamefont {Bichler}}, \bibinfo {author} {\bibfnamefont
  {G.}~\bibnamefont {Abstreiter}}, \bibinfo {author} {\bibfnamefont {M.~S.}\
  \bibnamefont {Brandt}}, \ and\ \bibinfo {author} {\bibfnamefont {J.~J.}\
  \bibnamefont {Finley}},\ }\href {\doibase 10.1063/1.3309684} {\bibfield
  {journal} {\bibinfo  {journal} {Applied Physics Letters}\ }\textbf {\bibinfo
  {volume} {96}},\ \bibinfo {pages} {053113} (\bibinfo {year}
  {2010})}\BibitemShut {NoStop}%
\bibitem [{\citenamefont {Toft}\ and\ \citenamefont {Phillips}(2007)}]{Toft07}%
  \BibitemOpen
  \bibfield  {author} {\bibinfo {author} {\bibfnamefont {I.}~\bibnamefont
  {Toft}}\ and\ \bibinfo {author} {\bibfnamefont {R.~T.}\ \bibnamefont
  {Phillips}},\ }\href {\doibase 10.1103/PhysRevB.76.033301} {\bibfield
  {journal} {\bibinfo  {journal} {Phys. Rev. B}\ }\textbf {\bibinfo {volume}
  {76}},\ \bibinfo {pages} {033301} (\bibinfo {year} {2007})}\BibitemShut
  {NoStop}%
\bibitem [{\citenamefont {Bayer}\ \emph {et~al.}(2002)\citenamefont {Bayer},
  \citenamefont {Ortner}, \citenamefont {Stern}, \citenamefont {Kuther},
  \citenamefont {Gorbunov}, \citenamefont {Forchel}, \citenamefont {Hawrylak},
  \citenamefont {Fafard}, \citenamefont {Hinzer}, \citenamefont {Reinecke},
  \citenamefont {Walck}, \citenamefont {Reithmaier}, \citenamefont {Klopf},\
  and\ \citenamefont {Sch\"afer}}]{Bayer}%
  \BibitemOpen
  \bibfield  {author} {\bibinfo {author} {\bibfnamefont {M.}~\bibnamefont
  {Bayer}}, \bibinfo {author} {\bibfnamefont {G.}~\bibnamefont {Ortner}},
  \bibinfo {author} {\bibfnamefont {O.}~\bibnamefont {Stern}}, \bibinfo
  {author} {\bibfnamefont {A.}~\bibnamefont {Kuther}}, \bibinfo {author}
  {\bibfnamefont {A.~A.}\ \bibnamefont {Gorbunov}}, \bibinfo {author}
  {\bibfnamefont {A.}~\bibnamefont {Forchel}}, \bibinfo {author} {\bibfnamefont
  {P.}~\bibnamefont {Hawrylak}}, \bibinfo {author} {\bibfnamefont
  {S.}~\bibnamefont {Fafard}}, \bibinfo {author} {\bibfnamefont
  {K.}~\bibnamefont {Hinzer}}, \bibinfo {author} {\bibfnamefont {T.~L.}\
  \bibnamefont {Reinecke}}, \bibinfo {author} {\bibfnamefont {S.~N.}\
  \bibnamefont {Walck}}, \bibinfo {author} {\bibfnamefont {J.~P.}\ \bibnamefont
  {Reithmaier}}, \bibinfo {author} {\bibfnamefont {F.}~\bibnamefont {Klopf}}, \
  and\ \bibinfo {author} {\bibfnamefont {F.}~\bibnamefont {Sch\"afer}},\ }\href
  {\doibase 10.1103/PhysRevB.65.195315} {\bibfield  {journal} {\bibinfo
  {journal} {Phys. Rev. B}\ }\textbf {\bibinfo {volume} {65}},\ \bibinfo
  {pages} {195315} (\bibinfo {year} {2002})}\BibitemShut {NoStop}%
\bibitem [{\citenamefont {van Kesteren}\ \emph {et~al.}(1990)\citenamefont {van
  Kesteren}, \citenamefont {Cosman}, \citenamefont {van~der Poel},\ and\
  \citenamefont {Foxon}}]{Kesteren90}%
  \BibitemOpen
  \bibfield  {author} {\bibinfo {author} {\bibfnamefont {H.~W.}\ \bibnamefont
  {van Kesteren}}, \bibinfo {author} {\bibfnamefont {E.~C.}\ \bibnamefont
  {Cosman}}, \bibinfo {author} {\bibfnamefont {W.~A. J.~A.}\ \bibnamefont
  {van~der Poel}}, \ and\ \bibinfo {author} {\bibfnamefont {C.~T.}\
  \bibnamefont {Foxon}},\ }\href {\doibase 10.1103/PhysRevB.41.5283} {\bibfield
   {journal} {\bibinfo  {journal} {Physical Review B}\ }\textbf {\bibinfo
  {volume} {41}},\ \bibinfo {pages} {5283} (\bibinfo {year}
  {1990})}\BibitemShut {NoStop}%
\bibitem [{\citenamefont {van Weert}\ \emph {et~al.}(2009)\citenamefont {van
  Weert}, \citenamefont {Akopian}, \citenamefont {Perinetti}, \citenamefont
  {van Kouwen}, \citenamefont {Algra}, \citenamefont {Verheijen}, \citenamefont
  {Bakkers}, \citenamefont {Kouwenhoven},\ and\ \citenamefont
  {Zwiller}}]{NL09Weert}%
  \BibitemOpen
  \bibfield  {author} {\bibinfo {author} {\bibfnamefont {M.~H.~M.}\
  \bibnamefont {van Weert}}, \bibinfo {author} {\bibfnamefont {N.}~\bibnamefont
  {Akopian}}, \bibinfo {author} {\bibfnamefont {U.}~\bibnamefont {Perinetti}},
  \bibinfo {author} {\bibfnamefont {M.~P.}\ \bibnamefont {van Kouwen}},
  \bibinfo {author} {\bibfnamefont {R.~E.}\ \bibnamefont {Algra}}, \bibinfo
  {author} {\bibfnamefont {M.~A.}\ \bibnamefont {Verheijen}}, \bibinfo {author}
  {\bibfnamefont {E.~P. A.~M.}\ \bibnamefont {Bakkers}}, \bibinfo {author}
  {\bibfnamefont {L.~P.}\ \bibnamefont {Kouwenhoven}}, \ and\ \bibinfo {author}
  {\bibfnamefont {V.}~\bibnamefont {Zwiller}},\ }\href {\doibase
  10.1021/nl900250g} {\bibfield  {journal} {\bibinfo  {journal} {Nano Letters}\
  }\textbf {\bibinfo {volume} {9}},\ \bibinfo {pages} {1989} (\bibinfo {year}
  {2009})}\BibitemShut {NoStop}%
\bibitem [{\citenamefont {Singh}\ and\ \citenamefont
  {Bester}(2009)}]{Bester09}%
  \BibitemOpen
  \bibfield  {author} {\bibinfo {author} {\bibfnamefont {R.}~\bibnamefont
  {Singh}}\ and\ \bibinfo {author} {\bibfnamefont {G.}~\bibnamefont {Bester}},\
  }\href {\doibase 10.1103/PhysRevLett.103.063601} {\bibfield  {journal}
  {\bibinfo  {journal} {Phys. Rev. Lett.}\ }\textbf {\bibinfo {volume} {103}},\
  \bibinfo {pages} {063601} (\bibinfo {year} {2009})}\BibitemShut {NoStop}%
\bibitem [{\citenamefont {Marie}\ \emph {et~al.}(1999)\citenamefont {Marie},
  \citenamefont {Amand}, \citenamefont {Le~Jeune}, \citenamefont {Paillard},
  \citenamefont {Renucci}, \citenamefont {Golub}, \citenamefont {Dymnikov},\
  and\ \citenamefont {Ivchenko}}]{Marie99}%
  \BibitemOpen
  \bibfield  {author} {\bibinfo {author} {\bibfnamefont {X.}~\bibnamefont
  {Marie}}, \bibinfo {author} {\bibfnamefont {T.}~\bibnamefont {Amand}},
  \bibinfo {author} {\bibfnamefont {P.}~\bibnamefont {Le~Jeune}}, \bibinfo
  {author} {\bibfnamefont {M.}~\bibnamefont {Paillard}}, \bibinfo {author}
  {\bibfnamefont {P.}~\bibnamefont {Renucci}}, \bibinfo {author} {\bibfnamefont
  {L.~E.}\ \bibnamefont {Golub}}, \bibinfo {author} {\bibfnamefont {V.~D.}\
  \bibnamefont {Dymnikov}}, \ and\ \bibinfo {author} {\bibfnamefont {E.~L.}\
  \bibnamefont {Ivchenko}},\ }\href {\doibase 10.1103/PhysRevB.60.5811}
  {\bibfield  {journal} {\bibinfo  {journal} {Physical Review B}\ }\textbf
  {\bibinfo {volume} {60}},\ \bibinfo {pages} {5811} (\bibinfo {year}
  {1999})}\BibitemShut {NoStop}%
\bibitem [{\citenamefont {Glasberg}\ \emph {et~al.}(1999)\citenamefont
  {Glasberg}, \citenamefont {Shtrikman}, \citenamefont {Bar-Joseph},\ and\
  \citenamefont {Klipstein}}]{Glasberg99}%
  \BibitemOpen
  \bibfield  {author} {\bibinfo {author} {\bibfnamefont {S.}~\bibnamefont
  {Glasberg}}, \bibinfo {author} {\bibfnamefont {H.}~\bibnamefont {Shtrikman}},
  \bibinfo {author} {\bibfnamefont {I.}~\bibnamefont {Bar-Joseph}}, \ and\
  \bibinfo {author} {\bibfnamefont {P.~C.}\ \bibnamefont {Klipstein}},\ }\href
  {\doibase 10.1103/PhysRevB.60.R16295} {\bibfield  {journal} {\bibinfo
  {journal} {Phys. Rev. B}\ }\textbf {\bibinfo {volume} {60}},\ \bibinfo
  {pages} {R16295} (\bibinfo {year} {1999})}\BibitemShut {NoStop}%
\bibitem [{\citenamefont {Kiselev}\ \emph {et~al.}(2001)\citenamefont
  {Kiselev}, \citenamefont {Kim},\ and\ \citenamefont
  {Yablonovitch}}]{Kiselev01}%
  \BibitemOpen
  \bibfield  {author} {\bibinfo {author} {\bibfnamefont {A.~A.}\ \bibnamefont
  {Kiselev}}, \bibinfo {author} {\bibfnamefont {K.~W.}\ \bibnamefont {Kim}}, \
  and\ \bibinfo {author} {\bibfnamefont {E.}~\bibnamefont {Yablonovitch}},\
  }\href {\doibase 10.1103/PhysRevB.64.125303} {\bibfield  {journal} {\bibinfo
  {journal} {Phys. Rev. B}\ }\textbf {\bibinfo {volume} {64}},\ \bibinfo
  {pages} {125303} (\bibinfo {year} {2001})}\BibitemShut {NoStop}%
\bibitem [{\citenamefont {Dymnikov}\ and\ \citenamefont
  {Konstantinov}(2009)}]{Dymnikov09}%
  \BibitemOpen
  \bibfield  {author} {\bibinfo {author} {\bibfnamefont {V.}~\bibnamefont
  {Dymnikov}}\ and\ \bibinfo {author} {\bibfnamefont {O.}~\bibnamefont
  {Konstantinov}},\ }\href {http://dx.doi.org/10.1134/S1063783409050023}
  {\bibfield  {journal} {\bibinfo  {journal} {Physics of the Solid State}\
  }\textbf {\bibinfo {volume} {51}},\ \bibinfo {pages} {884} (\bibinfo {year}
  {2009})},\ \bibinfo {note} {10.1134/S1063783409050023}\BibitemShut {NoStop}%
\bibitem [{\citenamefont {Kroutvar}\ \emph {et~al.}(2004)\citenamefont
  {Kroutvar}, \citenamefont {Ducommun}, \citenamefont {Heiss}, \citenamefont
  {Bichler}, \citenamefont {Schuh}, \citenamefont {Abstreiter},\ and\
  \citenamefont {Finley}}]{Nature04Finley}%
  \BibitemOpen
  \bibfield  {author} {\bibinfo {author} {\bibfnamefont {M.}~\bibnamefont
  {Kroutvar}}, \bibinfo {author} {\bibfnamefont {Y.}~\bibnamefont {Ducommun}},
  \bibinfo {author} {\bibfnamefont {D.}~\bibnamefont {Heiss}}, \bibinfo
  {author} {\bibfnamefont {M.}~\bibnamefont {Bichler}}, \bibinfo {author}
  {\bibfnamefont {D.}~\bibnamefont {Schuh}}, \bibinfo {author} {\bibfnamefont
  {G.}~\bibnamefont {Abstreiter}}, \ and\ \bibinfo {author} {\bibfnamefont
  {J.~J.}\ \bibnamefont {Finley}},\ }\href {\doibase 10.1038/nature03008}
  {\bibfield  {journal} {\bibinfo  {journal} {Nature}\ }\textbf {\bibinfo
  {volume} {432}},\ \bibinfo {pages} {81} (\bibinfo {year} {2004})}\BibitemShut
  {NoStop}%
\bibitem [{\citenamefont {Gupta}\ \emph {et~al.}(2001)\citenamefont {Gupta},
  \citenamefont {Knobel}, \citenamefont {Samarth},\ and\ \citenamefont
  {Awschalom}}]{Gupta01Science}%
  \BibitemOpen
  \bibfield  {author} {\bibinfo {author} {\bibfnamefont {J.~A.}\ \bibnamefont
  {Gupta}}, \bibinfo {author} {\bibfnamefont {R.}~\bibnamefont {Knobel}},
  \bibinfo {author} {\bibfnamefont {N.}~\bibnamefont {Samarth}}, \ and\
  \bibinfo {author} {\bibfnamefont {D.~D.}\ \bibnamefont {Awschalom}},\ }\href
  {\doibase 10.1126/science.1061169} {\bibfield  {journal} {\bibinfo  {journal}
  {Science}\ }\textbf {\bibinfo {volume} {292}},\ \bibinfo {pages} {2458}
  (\bibinfo {year} {2001})}\BibitemShut {NoStop}%
\bibitem [{\citenamefont {Mikkelsen}\ \emph {et~al.}(2007)\citenamefont
  {Mikkelsen}, \citenamefont {Berezovsky}, \citenamefont {Stoltz},
  \citenamefont {Coldren},\ and\ \citenamefont
  {Awschalom}}]{Mikkelsen07Nature}%
  \BibitemOpen
  \bibfield  {author} {\bibinfo {author} {\bibfnamefont {M.~H.}\ \bibnamefont
  {Mikkelsen}}, \bibinfo {author} {\bibfnamefont {J.}~\bibnamefont
  {Berezovsky}}, \bibinfo {author} {\bibfnamefont {N.~G.}\ \bibnamefont
  {Stoltz}}, \bibinfo {author} {\bibfnamefont {L.~A.}\ \bibnamefont {Coldren}},
  \ and\ \bibinfo {author} {\bibfnamefont {D.~D.}\ \bibnamefont {Awschalom}},\
  }\href {\doibase 10.1038/nphys736} {\bibfield  {journal} {\bibinfo  {journal}
  {Nature Physics}\ }\textbf {\bibinfo {volume} {3}},\ \bibinfo {pages} {770}
  (\bibinfo {year} {2007})}\BibitemShut {NoStop}%
\bibitem [{\citenamefont {Berezovsky}\ \emph {et~al.}(2008)\citenamefont
  {Berezovsky}, \citenamefont {Mikkelsen}, \citenamefont {Stoltz},
  \citenamefont {Coldren},\ and\ \citenamefont
  {Awschalom}}]{Berezovsky08Science}%
  \BibitemOpen
  \bibfield  {author} {\bibinfo {author} {\bibfnamefont {J.}~\bibnamefont
  {Berezovsky}}, \bibinfo {author} {\bibfnamefont {M.~H.}\ \bibnamefont
  {Mikkelsen}}, \bibinfo {author} {\bibfnamefont {N.~G.}\ \bibnamefont
  {Stoltz}}, \bibinfo {author} {\bibfnamefont {L.~A.}\ \bibnamefont {Coldren}},
  \ and\ \bibinfo {author} {\bibfnamefont {D.~D.}\ \bibnamefont {Awschalom}},\
  }\href {\doibase 10.1126/science.1154798} {\bibfield  {journal} {\bibinfo
  {journal} {Science}\ }\textbf {\bibinfo {volume} {320}},\ \bibinfo {pages}
  {349} (\bibinfo {year} {2008})}\BibitemShut {NoStop}%
\bibitem [{\citenamefont {Kosaka}\ \emph {et~al.}(2008)\citenamefont {Kosaka},
  \citenamefont {Shigyou}, \citenamefont {Mitsumori}, \citenamefont {Rikitake},
  \citenamefont {Imamura}, \citenamefont {Kutsuwa}, \citenamefont {Arai},\ and\
  \citenamefont {Edamatsu}}]{Kosaka08}%
  \BibitemOpen
  \bibfield  {author} {\bibinfo {author} {\bibfnamefont {H.}~\bibnamefont
  {Kosaka}}, \bibinfo {author} {\bibfnamefont {H.}~\bibnamefont {Shigyou}},
  \bibinfo {author} {\bibfnamefont {Y.}~\bibnamefont {Mitsumori}}, \bibinfo
  {author} {\bibfnamefont {Y.}~\bibnamefont {Rikitake}}, \bibinfo {author}
  {\bibfnamefont {H.}~\bibnamefont {Imamura}}, \bibinfo {author} {\bibfnamefont
  {T.}~\bibnamefont {Kutsuwa}}, \bibinfo {author} {\bibfnamefont
  {K.}~\bibnamefont {Arai}}, \ and\ \bibinfo {author} {\bibfnamefont
  {K.}~\bibnamefont {Edamatsu}},\ }\href {\doibase
  10.1103/PhysRevLett.100.096602} {\bibfield  {journal} {\bibinfo  {journal}
  {Phys. Rev. Lett.}\ }\textbf {\bibinfo {volume} {100}},\ \bibinfo {pages}
  {096602} (\bibinfo {year} {2008})}\BibitemShut {NoStop}%
\end{thebibliography}
\end{document}